\documentstyle [11pt,epsf,newpasp,twoside]{article}
\def\edcomment#1{\iffalse\marginpar{\raggedright\sl#1\/}\else\relax\fi}
\reversemarginpar

\def\ecs{\rm \,erg~ cm^{-2}\,s^{-1} }

\begin{document}

\title{The EMSS Radio Loud Quasar Sample}

\author{Anna Wolter}
\affil{Osservatorio Astronomico di Brera, via Brera 28, 20121
Milano, Italy}
\author{Annalisa Celotti} 
\affil{Sissa, via Beirut 2-4, 34014 Trieste, Italy}
\begin{abstract} 
We construct the first X-ray selected sample of radio-loud
quasars from the EMSS survey.  The X--ray selection allows us to
explore the properties of radio--loud quasars 10--100 
weaker than classical samples in the radio band.  There are no radio--loud
quasars whose synchrotron peak reaches the UV--soft X--ray band at
these (radio) flux levels (as occurs for BL Lac objects selected in
the X--ray band), but they appear to have comparatively stronger
optical--UV emission.  We suggest that this can be ascribed to a
significant contribution from a quasi--thermal optical--UV component,
emerging due to the comparatively weak non-thermal emission.  The
detection of sources at low radio fluxes also shows the presence of a
large population of steep spectrum quasars, and the lack of the
predicted turnover in the quasar counts.

\end{abstract}

\section{Introduction}

To understand a population properties it is crucial
to disentangle the intrinsic features from the selection induced ones.
To this aim it is necessary to compare the characteristics of sources 
selected in different spectral bands.
We have therefore considered radio--loud quasars -- which so far have 
been extracted from radio surveys -- and studied an X--ray selected sample.
We are mainly interested in determining whether the relation between 
the Spectral Energy Distribution (SED) and the source power proposed by 
Ghisellini et al. (1998) and Fossati et al. (1998) -- a blazar sequence from 
High-peaked BL Lacs (HBL), to Low-peaked BL Lacs (LBL) to Flat Spectrum Radio 
quasars (FSRQ) -- 
holds also for fainter X-ray selected quasars. The trend in the 
SED can be physically accounted for by an increase in the external 
radiation field along the sequence.

\section{The sample}

The sample under study comprises the radio--loud quasars detected in the
{\it Einstein} Medium Sensitivity Survey (EMSS; Gioia et al. 1990; 
Stocke et al. 1991).
It is the first statistically complete X--ray selected
sample of (39) radio--loud quasars. Details can be found in Wolter \&
Celotti (2000, A\&A submitted).  The sample is cut at $\delta \geq -40^{\circ}$, 
and excludes narrow line objects.  We dub the sample the EMSS 
Radio Loud quasar sample (ERL).
We divide the sample in flat (FS) and steep (SS) spectrum
radio--loud objects ($F_{\nu} \sim \nu^{-\alpha}$). The distribution
of $\alpha_r$ is not bimodal and therefore the choice of a dividing
value is somewhat arbitrary.  We formally consider it at
$\alpha_{r}$=0.7.

\section{Broad band properties}
\noindent
We study the 
luminosities in the radio, optical and X--ray bands ($L_{r}$, $L_{o}$,
$L_{x}$) and the corresponding broad band spectral indices
($\alpha_{ro}$, $\alpha_{ox}$, $\alpha_{rx}$).
Through X--ray selection we detect objects with a rather
limited range in optical luminosity, but a large span in the radio
one, extending the range of sampled $L_{r}$ towards 
fainter sources
compared to radio--selected quasars.

The radio and X-ray luminosities seem to be correlated at the
99\% level even taking into account the common redshift dependence.
No differences are found between the distributions of FS and
SS, neither in luminosities (KS probability $p\geq 20\%$) nor in broad
band spectral indices ($p \geq 47\%$) nor in the trends, suggesting a
behavior of the compact beamed component independent 
of the large
scale radio flux dominance.

The shape of the SED appears to be related to the radio
luminosity: in fact, the only statistically significant trends found
are between $\alpha_{ro}$, $\alpha_{rx}$ and $L_{r}$, consistent with
being caused by a change in the synchrotron peak energy: the
flatter is $\alpha_{ro}$, the higher is the peak energy.

\vskip -0.1cm
\begin{figure}
\plottwo{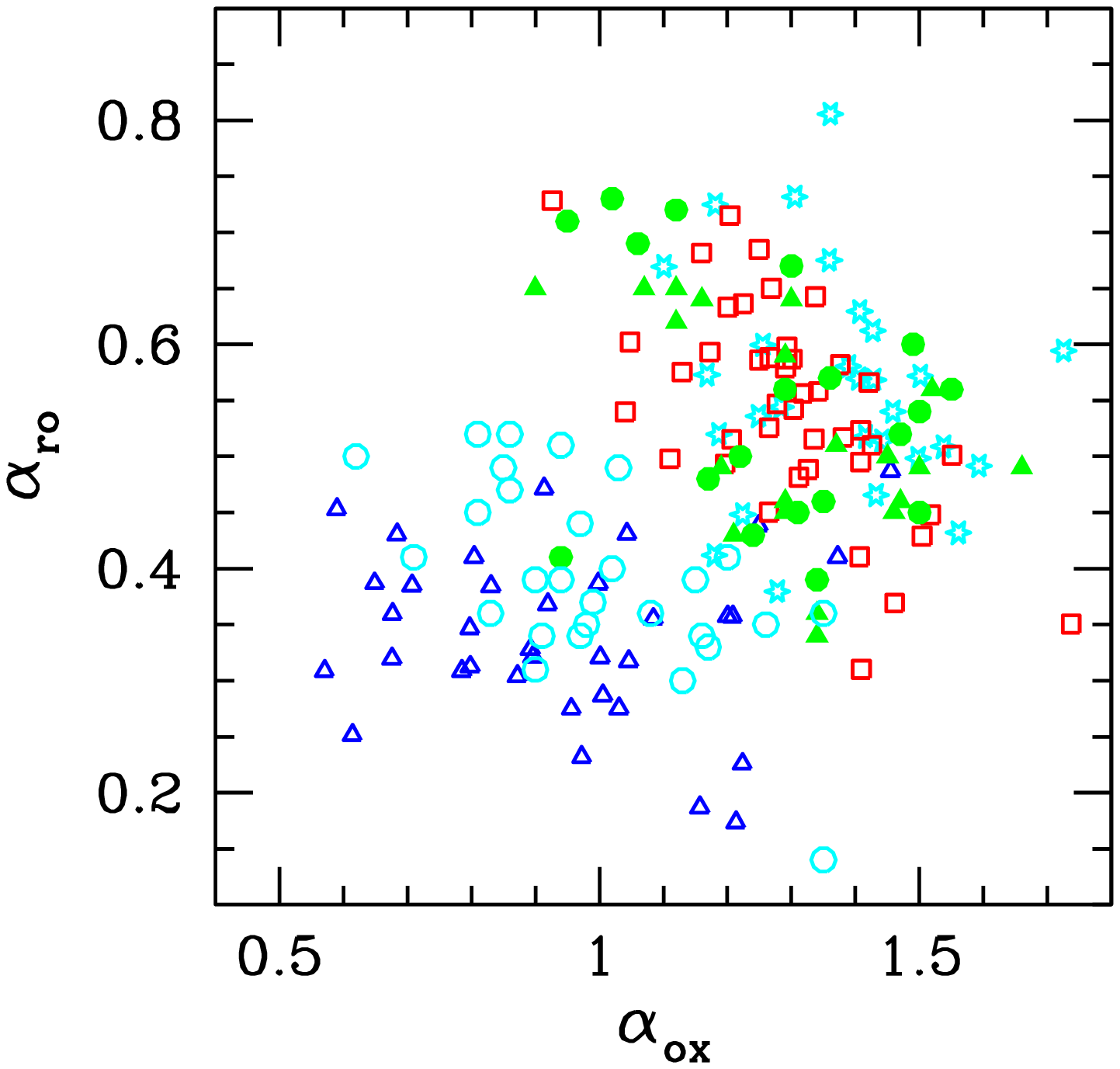}{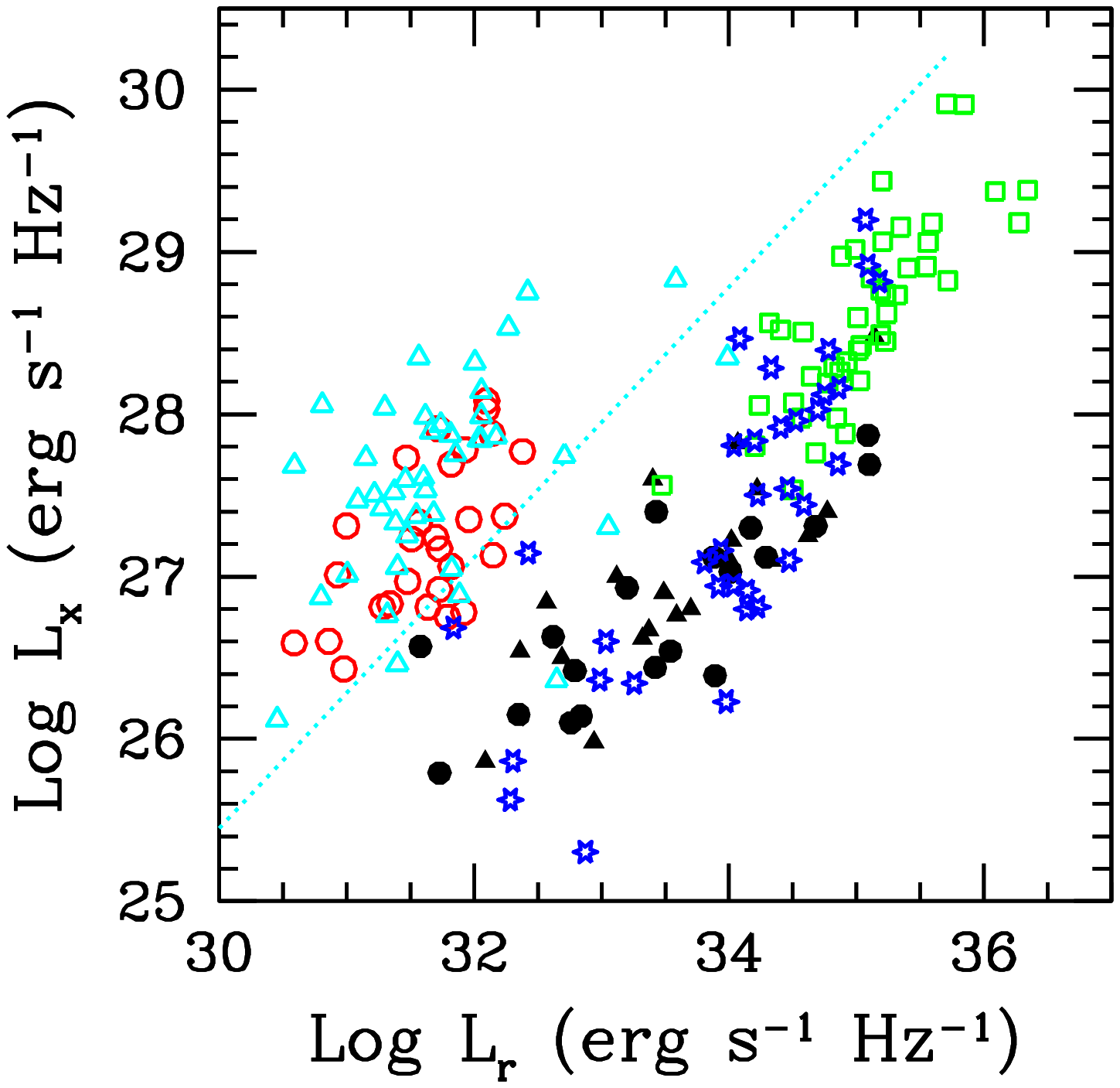}
\vskip -0.2cm
\caption{{\it Left:} Broad band spectral index $\alpha_{ro}$ vs. 
$\alpha_{ox}$ for
the ERL (filled triangles and circles for FS and SS, respectively)
compared to: EMSS BL Lacs (empty circles), Slew BL Lacs (empty
triangles), 1Jy BL Lacs (empty stars), 2Jy FSRQ (empty squares).
Just for graphical purposes, blazars selected by Perlman et al. (1998) and
Laurent-Muehleisen et al. (1999) are not included.
{\it Right:} X--ray vs radio luminosity for the same samples. 
The dotted line indicates the
``nominal separation'' between HBL and LBL in the luminosity plane (e.g. 
Fossati et al. 1998).
The ERL source in the HBL region is MS0815.7+5233, a weak lined AGN,
defined as ``BL Lac-like" in Stocke et al. (1991).
}
\label{alp_all}
\end{figure}

\section{Comparison with other samples}
 
\noindent
In Fig. 1a 
we show the ERL and samples of both FSRQ and
radio and X--ray selected BL Lacs.  The ERL occupy the transition
region (in the ``boomerang'' shaped blazar distribution) between FSRQ
and X--ray selected ($\sim$ HBL) BL Lac objects, roughly overlapping
with radio--selected ($\sim$ LBL) ones.  Note that the location of LBL
and FSRQ in the spectral index plane is due to the dominance in the
X--ray band of a flat Compton component.

Even the X--ray selection is unable to detect radio--loud
quasars with high peaked synchrotron components, albeit sampling sources with
$L_r$ comparable to that of HBL (see Fig. 1b). 
These findings are therefore in global agreement with the
expectation of the blazars sequence scenario, thus re-enforcing the
view of a strong connection between the SED and the radio luminosity.

We propose that the concentration of ERL in the LBL region
could be at least in part ascribed to the increasing 
dominance of
a quasi--thermal optically--UV component (blue bump) in quasars of
increasingly lower (non--thermal) radio power. Indeed the typical
$\alpha_{ro}$ and $\alpha_{rx}$ of the low radio power ERL 
corresponds to SED whose peak is located in the
optical--UV region.  The presence of such a component would
flatten $\alpha_{ro}$ and steepen $\alpha_{rx}$ in sources with
decreasing non--thermal continuum. Optical and X--ray spectra
will be able to verify such hypothesis.

\section{LogN-logS, Evolution, and Luminosity Functions}

\begin{figure}
\plottwo{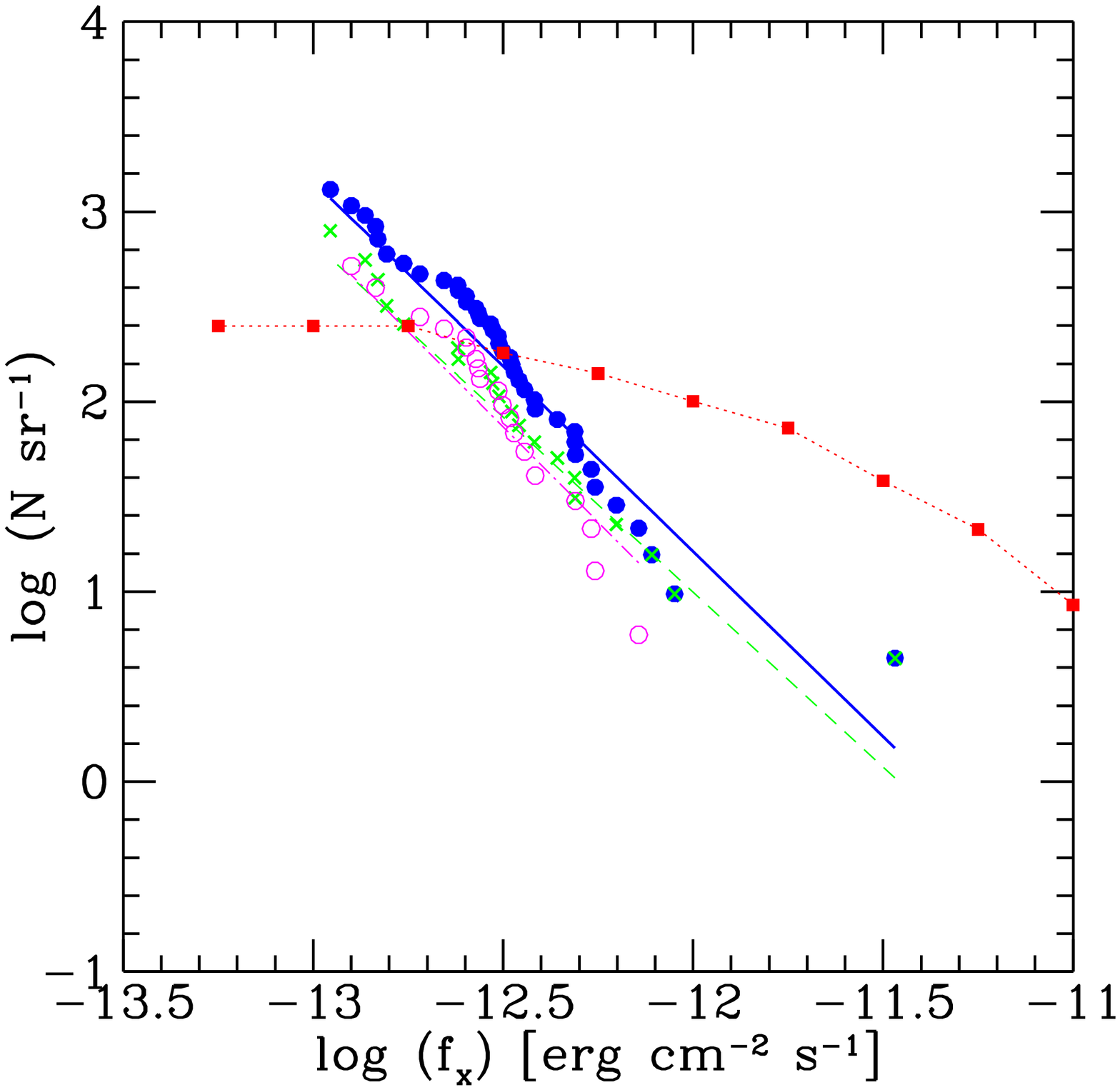}{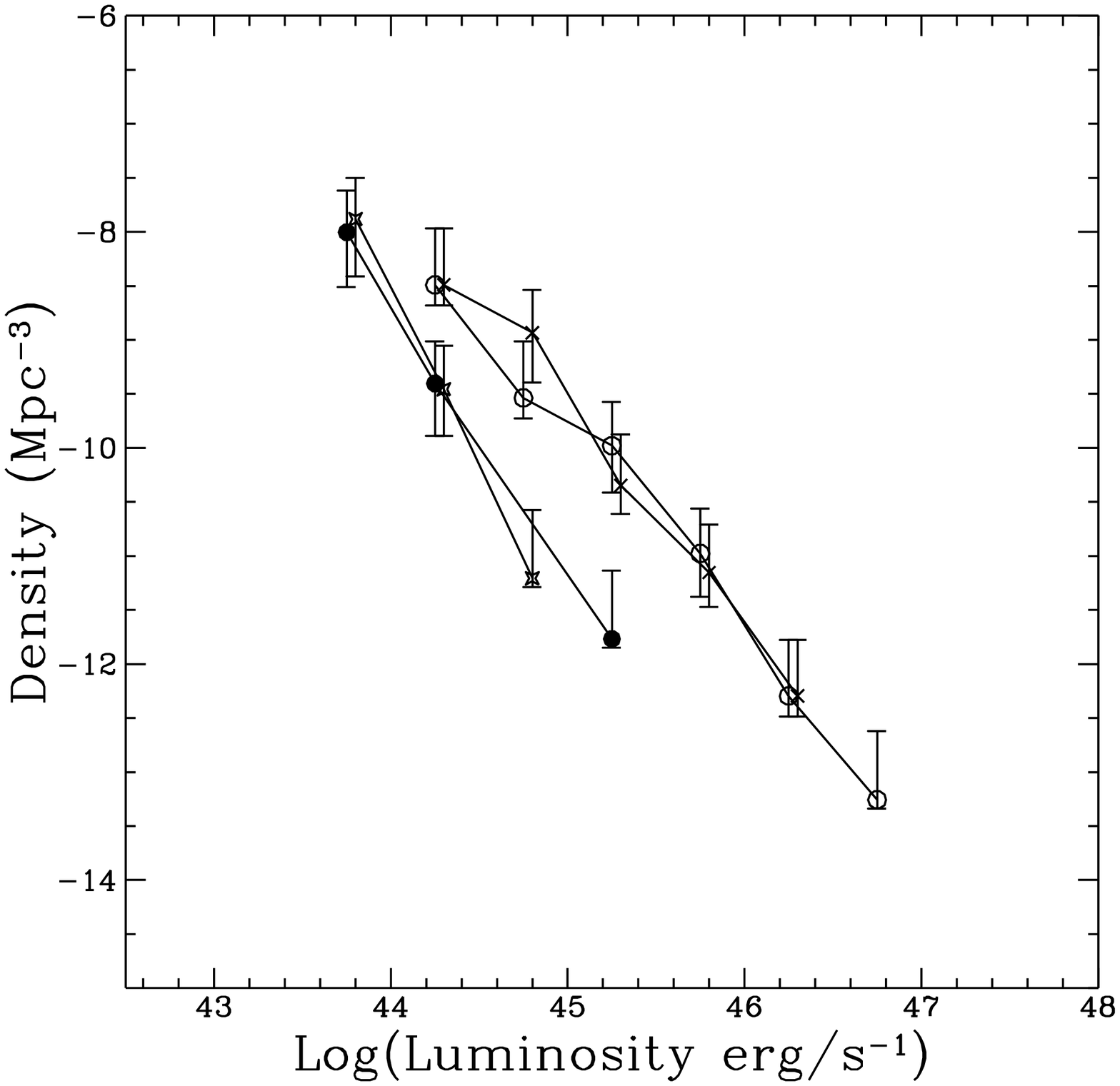}
\caption{{\it Left:} Integral number counts for FS (crosses), SS (circles) and all
ERL (filled circles). The lines represent the corresponding linear
fits: heavy line for ERL; dashed line for FS, and dot-dashed line for
SS. For comparison, the number counts of EMSS BL Lacs are reported
(from Wolter et al. 1991).
{\it Right:} Differential luminosity functions for the SS and FS sources,
in step of 0.5 $\log L$ (symbols as before; filled
circles and stars are de-evolved --exponential form-- SS and FS sources
respectively). SS points are shifted on the 
L axis for clarity.}
\label{figlnls}
\end{figure}

\vskip -0.8cm
\begin{table*}
\begin{center}
\caption{Statistical properties (in brackets the 1 $\sigma$ 
confidence range)}

\begin{tabular}[h]{| l c c c  |}
\hline

      & ERL&  FS & SS \\

\hline

Count slope &1.9 [1.7-2.2]  & 1.8 [1.6-2.1] & 2.0 [1.7-2.3] \\

$\langle$ z $\rangle$ & 0.92 $\pm$ 0.40& 0.99 $\pm$ 0.40& 0.85 $\pm$ 0.39\\

$V_e/V_a$    & 0.78 $\pm$ 0.05 & 0.76 $\pm$ 0.06  & 0.81 $\pm$ 0.07 \\

C &   & 6.6 [5.6-7.4]  & 6.4 [5.6-7.1] \\


LF integral slope & 1.6 [1.4-1.7] & 1.6 [1.3-1.8] & 1.9 [1.6-2.2]\\

\hline
\end{tabular}
\end{center}
\label{stat}
\end{table*}

\noindent
We plot the number counts of FS and SS sources in
Fig. 2a. 
There is marginal evidence
(though over the whole range in flux) for flatter 
counts of FS (see Table 1.) 
The count distributions of FS is in complete agreement with what found
for all AGN -- mostly radio quiet-- in the EMSS ($1.61 \pm 0.06$,
Della Ceca et al. 1992); SS and all ERL counts are marginally 
steeper
(at 1 $\sigma$ level) than all AGN, possibly indicating larger
amount of evolution.

The most interesting aspect is the extension to a factor 100
lower fluxes provided by the X--ray selection.
At radio fluxes of a few tens of mJy (which corresponds for an average
SED to $\sim 10^{-13} \ecs $)
the radio counts of X--ray selected objects are marginally
consistent with the Euclidean extrapolation from higher fluxes and
thus largely exceed (factor $>$10 for FS and $\sim$3 for SS)
those predicted by the beaming model, which drop below $\sim$
0.1 Jy (Padovani \& Urry 1992).
The EMSS BL Lac distribution (see Fig. 2a) 
is flatter, dominating by a factor $\sim 10$ at $\sim 10^{-12} \ecs$,
and an `inversion' of the two populations occurs just above
$\sim$3 $\times 10^{-13} \ecs$.

We have studied the cosmological evolution by applying the
V$_e$/V$_a$ test (Avni \& Bachall, 1980) - see Table 1. 
The derived evolution rate for ERL and FS is consistent with the total 
EMSS AGN sample (Della Ceca et al. 1992).  SS require a slightly higher
value for the evolution parameter.
The differential luminosity functions are plotted in Fig. 2b 
as observed, and de-evolved at z=0.
Derived slopes are in  Table 1. 

\begin{acknowledgements}
\noindent
This work has received partial finantial support from 
the Italian MURST under grant
COFIN98-02-32. 
\end{acknowledgements}

\end{document}